# Intense femtosecond photoexcitation of bulk and monolayer MoS$_2$

I. Paradisanos[1,2], E. Kymakis[3], C. Fotakis[1,2], G. Kioseoglou[4], E. Stratakis[1,4,a)]

[1] *Institute of Electronic Structure and Laser (IESL), Foundation for Research and Technology-Hellas (FORTH), Heraklion, 71003, Greece*

[2]*Physics Department, University of Crete, Heraklion, 71003, Greece*

[3]*Center of Materials Technology and Photonics & Electrical Engineering Department, Technological Educational Institute (TEI) of Crete, Heraklion, 71003, Greece*

[4]*Materials Science and Technology Department, University of Crete, Heraklion, 71003, Greece*

## ABSTRACT

The effect of femtosecond laser irradiation on bulk and single-layer MoS$_2$ on silicon oxide is studied. Optical, Field Emission Scanning Electron Microscopy (FESEM) and Raman microscopies were used to quantify the damage. The intensity of $A_{1g}$ and $E_{2g}^1$ vibrational modes was recorded as a function of the number of irradiation pulses. The observed behavior was attributed to laser-induced bond breaking and subsequent atoms removal due to electronic excitations. The single-pulse optical damage threshold was determined for the monolayer and bulk under 800nm and 1030nm pulsed laser irradiation and the role of two-photon versus one photon absorption effects is discussed.

a) Corresponding author: stratak@iesl.forth.gr

Atomically thin two-dimensional (2D) materials including graphene have attracted significant research interest due to their extraordinary physical properties.[1,2] However, graphene is a zero band gap material which in some cases is an undesirable property for optoelectronic applications.[3] Transition metal dichalcogenides (TMDs) combine the 2D layered structure with a finite band gap and therefore are considered to be alternatives of graphene.[4] Among TMDs, $MoS_2$ is one of the most stable ones and in bulk form it is an indirect band gap semiconductor with an energy gap, $E_g$, of ~ 1.3 eV. Upon reducing the number of layers $MoS_2$ undergoes a transition from an indirect to a direct band gap semiconductor reaching an optical gap of ~1.9 eV in monolayer form.[2,5-7] Apart from its direct bandgap, monolayer $MoS_2$ exhibits a stable charge exciton state even at room temperature, a property that is desirable for various optoelectronic and photonic applications[8], including phototransistors[9,10], light emitters[11], and heterojunction solar cells.[12,13]

Towards the development of 2D photonic devices, the investigation of 2D materials response under intense photoexcitation by ultrashort pulses, as well as of their ultrafast optical properties, including nonlinear susceptibility, refraction, absorption and carrier relaxation, is undoubtedly important. For example, recent investigations have focused on the non-linear optical properties of monolayer $MoS_2$[14] paving the way for applications including mode-lock devices, laser protection optical limiters, saturable absorbers and optical switches. Another aspect of the intense photoexcitation properties of monolayer $MoS_2$ is the evolution of the structure during heating, melting/resolidification and finally optical damage. Although, a previous work has shown that continuous wave (CW) laser irradiation can be employed for photo-thermal thinning of bulk $MoS_2$ crystals down to monolayer[15], to date there is no investigation of the $MoS_2$ lattice response under ultrashort pulsed laser photoexcitation conditions. A unique characteristic of ultrashort (i.e. sub-picosecond) laser-material interaction is that the photon energy is transferred to the lattice at rates faster than the electron-phonon relaxation time. Such ultrafast absorption process could give rise to both thermal and non-thermal effects within the lattice. In addition to this, the study of the monolayer response compared to that of the bulk is of great interest, considering the fundamental differences in optical absorption properties of direct compared to indirect gap semiconductors.

In this work, the effect of intense femtosecond laser excitation on the structure of bulk and monolayer $MoS_2$, under conditions ranging from lattice heating to material damage is systematically investigated. In particular, the evolution of Raman active vibrational modes in bulk whose thickness was ranging from 60 to 80 nm for several samples studied and monolayer $MoS_2$ was recorded as a function of irradiation intensity and total exposure time. Additionally, the single-pulse damage threshold in monolayer $MoS_2$ was identified and compared to that of the bulk crystal. Experiments reveal large differences in the ultrafast laser excitation response of monolayer compared to the bulk, as far as the lattice distortion as well as the lattice morphology at the onset of optical damage.

Few-layer $MoS_2$ samples were mechanically exfoliated from a bulk natural crystal (SPI Supplies) using an adhesive tape and subsequently deposited on Si/Silicon Oxide (290nm) wafers. Single layer regions of 3–5 μm across were identified with an optical microscope and confirmed with micro-Raman (Thermo Scientific) spectroscopy at room temperature. Photoluminescence (PL) studies from 80K to 300K were employed using a micro-PL setup in backscattering geometry and a 532nm continuous wave laser. The irradiation experiments were performed using a 200 fs Ti:Sapphire laser system operating at 800 nm wavelength and 1 kHz repetition rate. The energy of the beam was controlled via a combination of a waveplate, a linear polarizer and a series of neutral density filters, while the polarization direction was controlled via a λ/2 waveplate. An iris aperture was used to obtain the central part of the beam and acquire a uniform energy distribution. The laser beam was focused down to 100 μm on the sample, placed on an XYZ translation stage, at normal incidence. In a typical experiment the laser beam irradiated monolayers adjacent to bulk areas in an effort to compare the response of such different areas upon excitation. The alignment and irradiation processes could be continuously monitored by means of a charge-coupled device (CCD) imaging setup. Figure 1a shows a typical optical microscopy image of exfoliated flakes, where areas of different layer numbers are indicated. Micro-Raman spectroscopy, using a 473 nm excitation wavelength with very low intensity in order to avoid structural damage, was utilized to identify the number of layers in certain areas of the exfoliated flakes. The energy difference between the two most prominent Raman vibrational modes, $A_{1g}$ and $E_{2g}^1$, is used extensively in the literature as the fingerprint of the number of layers[16]; $A_{1g}$ is the out-of-plane and $E_{2g}^1$ is the in plane vibrational mode whose energy at the monolayer limit is 402 cm$^{-1}$ and 384 cm$^{-1}$ respectively. Typical Raman spectra of single layer, bilayer and bulk at T=300K are shown in Fig.2a. The low-energy shoulder is an artifact of the measurement apparatus since it is also observed in the Si Raman peak at 520 cm$^{-1}$. The energy difference of 18cm$^{-1}$ for the lower spectrum unambiguously confirms the single layer $MoS_2$. To further confirm the existence of monolayers we have performed temperature dependent photoluminescence spectroscopy. The inset of Fig. 2b shows a typical PL emission spectrum taken at 80K using a 532nm laser as an excitation source. The main peak of this spectrum is the strong PL emission at approximately 1.90 eV which is due to the direct A-exciton transition at the K-point of the Brillouin zone.[2,6,7] The temperature dependence of this feature, shown in Fig. 2b, exhibits a standard semiconductor-like behavior and can be described by the Varshni relation. The solid line is a fit to the data using the values α=0.33meV/K and β=238K for the Varshni parameters. These PL characteristics are in accordance with previous studies presented in the literature.[7,17]

      We initially investigated a single-pulse optical damage threshold of monolayer $MoS_2$ and compare it with that of the bulk. We define such threshold as the single-pulse light flux at which submicron-sized distortion features are created in either of the two cases. We observed that the damage threshold of the monolayer is very well-defined i.e. no modification is

observed up to a certain fluence of 50 mJ/cm$^2$ (2.5mW), beyond which damage occurs via material ablation. Interestingly, it is found to be higher than that of the bulk that is 15 mJ/cm$^2$ (750μW). These values correspond to time averaged power.

Figure 1b shows an optical microscopy image of MoS$_2$ bulk and monolayer flakes exposed to 10$^3$ pulses at a fluence of 20 mJ/cm$^2$ that is above the bulk damage threshold but below that of the monolayer. It is evident that while bulk areas were significantly distorted due to material ablation, the single layer was practically unaffected by laser irradiation. However, upon increasing the number of pulses to 10$^5$, while keeping the fluence at 20mJ/cm$^2$, the monolayer can also be affected. This is confirmed by the respective field emission scanning electron microscopy (FESEM) image of Fig. 1c, showing that the single layer is distorted as well. Raman spectroscopy was employed to investigate the characteristics of the lattice modifications induced in bulk and monolayer MoS$_2$ as a function of the irradiation intensity and total exposure time. A first observation was a significant decrease of the intensities of $A_{1g}$ and $E_{2g}^1$ Raman modes recorded from the ablated areas compared to the pristine ones while no energy shift of these peaks was observed. A similar effect has been observed during femtosecond laser interaction with graphene.[18,19] To further shed light on this effect we monitored the evolution of the Raman spectra as function of the number of irradiation pulses N (proportional to the exposure time), using laser fluences below the single-shot damage threshold. We first irradiated the sample with the femtosecond laser, then we performed Raman measurements in a different experimental setup and then we irradiated the sample again at the same spot, repeating this process several times with different number of pulses each time. Fig. 3a (3b) shows the dependence of the intensity for the out of plane $A_{1g}$ ($E_{2g}^1$) mode to the number of pulses N for the monolayer and bulk at 600μW. This laser power is below the damage threshold for both bulk and monolayer. The data were normalized to the single pulse intensity (N=1). In case of the monolayer, the $A_{1g}$ and $E_{2g}^1$ peak intensities were practically constant with N until they rapidly decreased at a critical exposure time. Such abrupt decrease possibly indicates ablation and eventual sublimation of the $MoS_2$ lattice. On the other hand, for bulk $MoS_2$, a monotonic decrease of the $A_{1g}$ and $E_{2g}$ peak intensities was observed for low N followed by a more abrupt decrease than that recorded for the single layer. Compared to the monolayer, the latter occurred at lower number of irradiation pulses. Both of the above results comply with the lower optical damage threshold observed for the bulk compared to the single layer. It should be noted that the above results were repeatable within experimental uncertainty over different probing sites within the bulk and/or monolayer areas. The inset of Fig. 3a (3b) shows Raman spectra for the $A_{1g}$ ($E_{2g}^1$) mode for monolayer and bulk taken under N=1 and N=$10^5$ irradiation pulses. The plots clearly demonstrate the dramatic effect on the intensity that a large number of pulses has on the bulk with respect to the monolayer. Again, the spectra were normalized to the single pulse intensity (N=1). Lorentzian fit analysis in our raw data showed that the linewidth of the two main Raman modes, as a function of the number of irradiation pulses, remains practically constant within the experimental error.

Figures 4a and 4b present the pulse number dependence of the Raman peaks for the monolayer at two different fluences (600μW and 1mW, both below the single-shot single-layer damage threshold). It is evident that the peak intensity of both Raman modes is a strong function of the irradiation intensity. We postulate that the observed weakening of the Raman peaks can be attributed to a corresponding decrease in the number of scattering centers (i.e. Mo-S bonds), possibly due to laser-induced bond breaking and subsequent atoms removal. Indeed, as *N* was increased, optical and FESEM microscopy revealed the formation of macroscopic holes within the flake area that progressively increased in size upon further exposure. This is in accordance to previous observations on radiation induced damage studies showing that as soon as the $MoS_2$ sheet is perforated by losing first a single S atom and subsequently the other Mo–bonded S atom via ionization, this hole readily enlarges.[20] It should be noted that the absence of a peak at 820 cm$^{-1}$ in the Raman spectra of the irradiated areas, which is the fingerprint associated with the high-temperature oxidation of $MoS_2$ to $MoO_3$, suggests that the $MoS_2$ surface is not affected by oxidation during the laser irradiation process. Therefore, it can be proposed that the damage of nanosheets occurs through lattice sublimation. Interestingly, we have observed that CW laser irradiation of natural crystal $MoS_2$ did not give rise to oxidation as well; this is in contrast to microcrystalline $MoS_2$ powder which is found to oxidize under irradiation at similar conditions.[21] For the ultrafast-matter interaction, electronic excitations should be significant in the timescale of hundreds of fs, due to the absence of electron-phonon coupling during the pulse. Under such conditions electrons are excited from bonding to anti-bonding states while the energy due to recombination of photoexcited carriers facilitates photochemical bond breaking. Investigation of the dependence of laser induced lattice degradation process on pulse duration may assist towards understanding the photothermal and photochemical contributions. For this purpose, experiments comparing fs with longer pulses are currently in progress.

Optical breakdown and subsequent lattice distortion occurs when photogenerated electron density in the conduction band reaches a critical value. The monolayer $MoS_2$ is a direct semiconductor with an optical gap of ~1.90 eV, while the bulk $MoS_2$ is an indirect semiconductor with an optical gap of ~1.30 eV. Considering the laser photon energy of 1.55 eV ($\lambda$= 800 nm) used in our case, single-photon absorption dominates for the multilayer $MoS_2$, while monolayer can only be excited via a two-photon absorption (TPA) process (Fig. 1d) that shows a quadratic dependence on the laser intensity. One consequence of TPA-governed laser-induced breakdown is an increase by more than three times of the damage threshold[22], since the absorption of two photons is by orders of magnitude less probable than the single-photon absorption process.[23] Therefore, in our case, the probability of reaching the critical electron excitation required for lattice decomposition is much lower in case of the larger-gap monolayer $MoS_2$. To further explore the above proposed mechanism, single-shot irradiation experiments were performed using a same repetition-rate laser source, with a photon energy of 1.20 eV (λ=1030 nm - Yb-doped Potassium-Gadolinium Tungstate crystal) which is below the gap of both the monolayer and the bulk. In this case the single-shot damage threshold of monolayer and bulk $MoS_2$ were found to be comparable, i.e 300 and 240 mJ/cm$^2$ respectively, as indeed anticipated due to the occurrence of TPA process in both cases. Unlike the small differences in the absorbance between 1.55eV and 1.2eV for monolayer $MoS_2$,[2] there is a six-time rise in the single-shot damage threshold. However, considering the non-linear nature of absorption process, small differences in the absorbance could correspond to a non-linear relation of the respective damage threshold differences.

Finally, it should be noted that the optical damage in bulk $MoS_2$ gives rise to interesting pseudo-periodic (period of ~200 nm) ripple-like patterns (Fig. 1c) oriented parallel to the laser polarization. Ripples' formation is commonly observed in semiconducting materials and can be mainly attributed to electromagnetic interference effects.[24] Formation of self-assembled surface structures is an interesting aspect of the interaction of ultrafast lasers with 2D materials. Experiments are on-going to investigate the physics behind the generation of the observed patterns in the case of $MoS_2$.

In summary, we have investigated the effect of intense femtosecond laser excitation on the structure of bulk and monolayer $MoS_2$. The evolution of $A_{1g}$ and $E_{2g}^1$ vibrational modes was recorded as a function of irradiation intensity and total exposure time. The observed behavior of the Raman peaks could be attributed to a corresponding decrease in the number of Mo-S bonds, possibly due to laser-induced bond breaking and subsequent atoms removal. Since no $MoO_3$ formation was observed in the Raman spectra, it can be proposed that the damage of nanosheets occurs through lattice sublimation. The single-pulse optical damage threshold was determined for the monolayer and bulk to be 50mJ/cm$^2$ and 15mJ/cm$^2$, respectively, under 800nm wavelength irradiation. The more than 3 times higher damage threshold for the monolayer is understood as the result of a two-photon absorption process versus single-photon absorption in the bulk. This mechanism was verified under a higher wavelength (1030 nm) irradiation where two-photon absorption is required not only for the single layer but also for the bulk.

**ACKNOWLEDGMENTS**

This work is partially supported by the Integrated Initiative of European Laser Research Infrastructures LASERLAB-II (Grant Agreement No. 228334).

**Figure Captions**

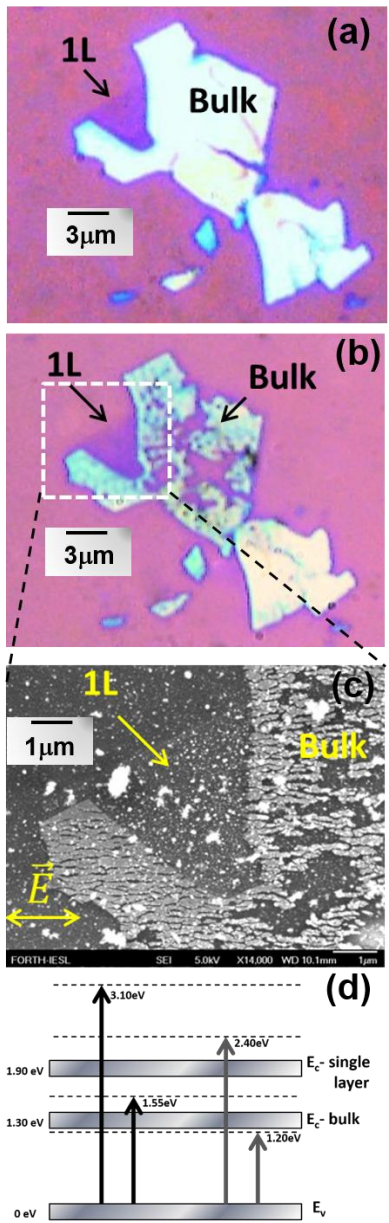

**Fig.1.** (a) Optical microscope image of pristine bulk and single layer $MoS_2$ obtained by mechanical exfoliation on 290nm Silicon oxide (b) Optical microscope image of the sample irradiated by 20 mJ/cm$^2$ and $10^3$-200fs-pulses of an 800nm-1kHz laser (c) FESEM image of the sample (dashed line square region in Fig1.b) irradiated by 20 mJ/cm$^2$ and $10^5$ pulses. Ripple formation parallel to the electric field vector on bulk $MoS_2$ and single layer distortion (d) Schematic representation showing the optical gap of bulk and monolayer $MoS_2$ and the electron excitation via single and two photon absorption (TPA) process of 800nm (1.55eV) and 1030nm (1.20eV).

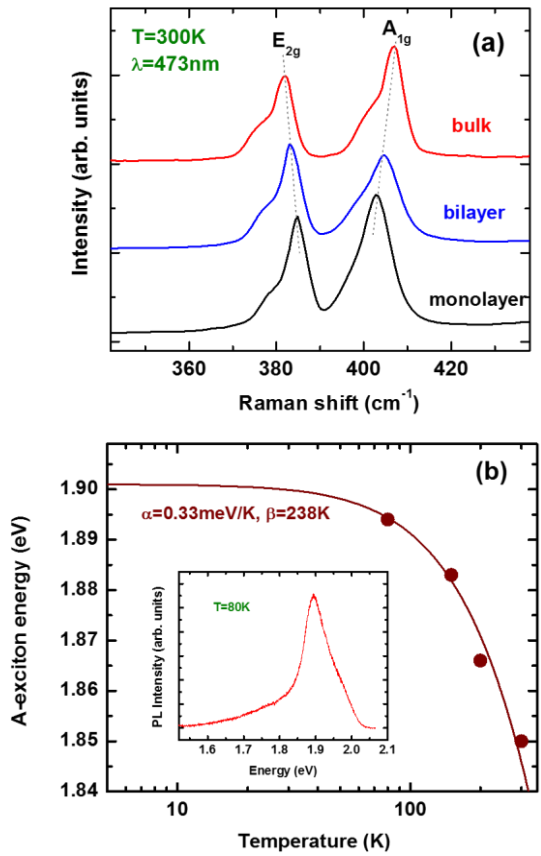

**Fig.2.** (a) Raman spectra at T=300K for bulk, bilayer and monolayer $MoS_2$. The energy separation between the in plane $E^1_{2g}$ and the out of plane $A_{1g}$ is 18 cm$^{-1}$, characteristic of single layer (b) Temperature dependence of the photoluminescence taken with 532 nm excitation shows a typical semiconductor behavior. The PL is dominated by the A-exciton emission (inset, PL at 80K).

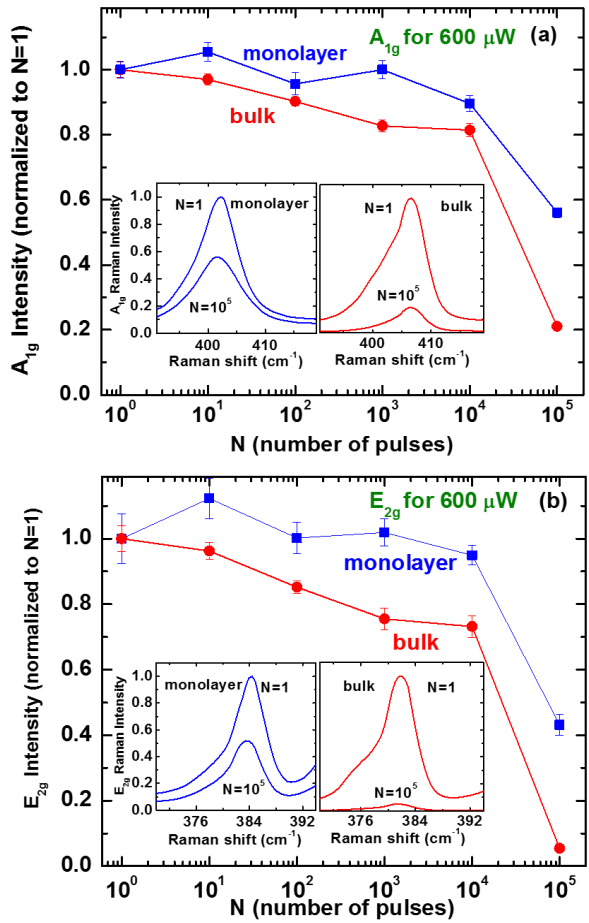

**Fig.3.** Intensity of the out of plane $A_{1g}$ (a) and in plane $E_{2g}^1$ mode (b) as function of the number of pulses N for the monolayer and bulk at 600μW (normalized to N=1). The inset shows the $A_{1g}$ (a) and $E_{2g}^1$ (b) Raman spectra for monolayer and bulk taken under N=1 and N=$10^5$ irradiation pulses.

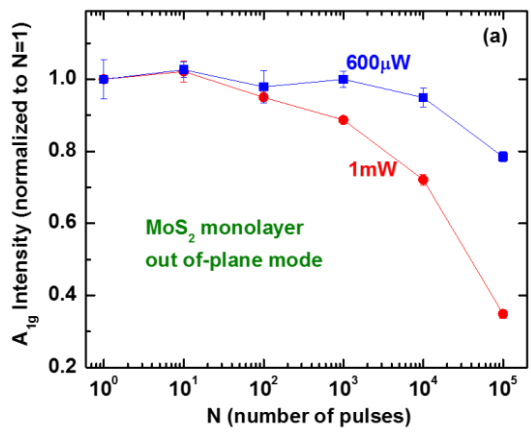

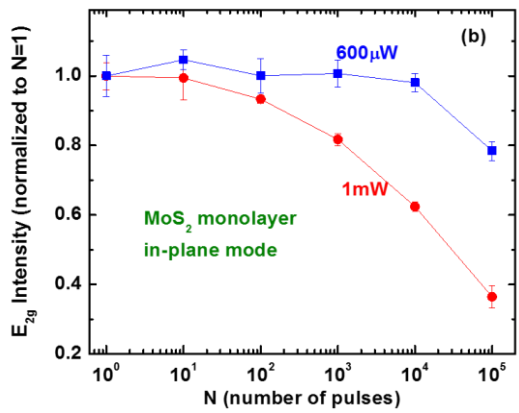

**Fig.4**. Normalized (to N=1) intensity of the out of plane $A_{1g}$ (a) and in plane $E_{2g}^1$ mode (b) for the single layer as function of the number of pulses N at two different fluences.